\documentclass[12pt]{revtex4}

\usepackage{natbib}
\usepackage{graphicx}

\begin{document}

\newcommand{\baseven}{$^{137}\rm{Ba}^{+}$}

\title{Hyperfine and Optical Barium Ion Qubits}

\author{M. R. Dietrich}
\email{dietricm@u.washington.edu}
\author{N. Kurz}
\author{T. Noel}
\author{G. Shu}
\author{B. B. Blinov}
\affiliation{University of Washington Department of Physics, Seattle, Washington 98195}

\begin{abstract} 

State preparation, qubit rotation, and high fidelity readout are
demonstrated for two separate \baseven~qubit types.  First, an optical qubit on
the narrow 6S$_{1/2}$ to 5D$_{5/2}$ transition at 1.76 $\mu$m is implemented.
Then, leveraging the techniques developed there for readout, a ground state
hyperfine qubit using the magnetically insensitive transition at 8 GHz is
accomplished.

\end{abstract}

\maketitle

\section{Introduction}

A quantum computer, once implemented, is expected to have several applications
with significant scientific and social utility, due to its ability to execute
specific algorithms with scaling laws quadratically or exponentially faster
than any known classical algorithm that accomplishes the same task
\cite{Nielsen}.  The fundamental unit of information in a quantum computer is
the qubit, which is physically realized by any coherent quantum two-level
system.  At present, the best developed qubit technology is provided by the
hyperfine or optical levels of an elemental ion suspended in a radio frequency
(RF) Paul trap \cite{Ladd2010} .  That fact is in no small part thanks to
decades of development of techniques in atomic physics to manipulate these
profoundly quantum systems.

\begin{figure}
\includegraphics[width=3in]{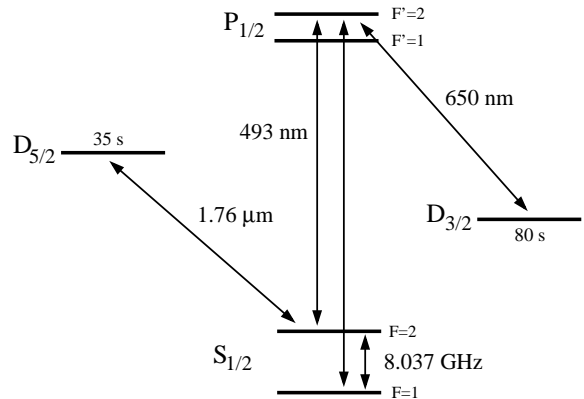}

\caption{A representation of the energy levels in \baseven, and the transitions
accessed by various lasers.  The ion is cooled on the 493 nm transition with
8.037 GHz sidebands and repumped with a 650 nm laser diode.  Shelving to the
D$_{5/2}$ state is accomplished through coherent excitation with the 1.76
$\mu$m laser.  Hyperfine structure of the D states is not shown.}

\label{leveldiagram}
\end{figure}

Despite a long history in ion trapping \cite{Dehmelt1980,Janik1985,DeVoe2002},
barium has never previously been demonstrated as an ionic qubit.  Ba$^{+}$ has
several desirable characteristics for an ionic qubit.  The odd isotope
\baseven~is relatively abundant at 11\%, which is sufficiently high that it can
be trapped from a natural source without isotope selective ionization
\cite{Dietrich2008,Steele2007}.  Barium possesses a long lived metastable state
$5^2$D$_{5/2}$ (see figure \ref{leveldiagram}), whose lifetime (35 s) is an
order of magnitude greater than that of any other singly ionized alkali earth
atom \cite{Madej1990}.  It is important for this state to be long lived, since
that decay rate restricts the qubit readout fidelity and sets a physical upper
limit to optical qubit coherence times.  Furthermore, the laser wavelengths for
the 6S$_{1/2}$ to 5D$_{5/2}$ infrared ``shelving'' transition as well as
barium's visible wavelength cooling transition are the longest of any ionic
qubit candidate, which makes it favorable for remote photonic coupling through
optical fiber.  This property is essential for long distance entanglement
between ionic qubits \cite{Simon2003,Olmschenk2009}, since the post selection
entanglement process is mediated by emitted photons \cite{Moehring2004}.
Finally, barium atomic structure is well understood, since it has been studied
extensively both theoretically \cite{Gopakumar2002,IskrenovaTchoukova2008} and
experimentally for possible applications as an optical frequency standard
\cite{Sherman2005,Madej1992} and for a test of parity non-conservation
\cite{Fortson1993,Sherman2005b}. 

Here we report the trapping, state preparation, state rotation, and readout of
two different \baseven~ionic qubits, one based on the narrow infrared optical
transition at 1.76 $\mu$m and the other based on the ground state hyperfine
splitting.  In the former case, sufficient coherence is observed on the
quadrupole transition in a Doppler-cooled ion to drive about 10 Rabi rotations.
Since the excited D$_{5/2}$ state is disjoint from the laser cooling cycle,
readout is effected by simply enabling the cooling lasers and looking for
fluorescence.  In the latter case, qubit rotations are driven by a resonant
microwave radiation pulse near 8 GHz and readout is achieved directly with Rabi
oscillation on the 1.76 $\mu$m transition to selectively shelve the ion
into the metastable state.

\section{Experimental Setup}

The energy level diagram for \baseven~can be found in figure
\ref{leveldiagram}.  The ion is Doppler cooled on the D1 line at 493 nm using a
frequency-doubled 986 nm external cavity diode laser (ECDL).  Since this state
decays into the D$_{3/2}$ state about 25\% of the time, the ion must then be
repumped out using a second ECDL at 650 nm.  \baseven~has a nuclear spin of
I=3/2; in order to optically address all hyperfine levels in the S and D
states, the 493 nm laser beam is modulated with an electro-optic modulator
(EOM) at 8.037 GHz, while the 650 nm laser is directly current modulated at 614,
539 and 394 MHz using a bias-T.  The 493 nm laser carrier is set to the
$F=1\rightarrow F'=2$ transition while the sideband drives $F=2\rightarrow
F'=2$, where the prime indicates the P$_{1/2}$~manifold.  This configuration
avoids the weak $F=1\rightarrow F'=1$ transition and allows us to optically
pump into the $F=2,m_F=0$ state using $\pi$ polarized light.  For optimal
cooling, the 493 nm laser is set to an elliptical polarization.

The 986 nm laser is frequency stabilized using an Invar-spaced optical cavity
with a free spectral range of 1.2 GHz and finesse of about 150, which is in a
sealed, temperature controlled cylinder.  A frequency shift is provided by a
double-passed acousto-optic modulator (AOM), whose feed RF is frequency
modulated.  We then detect the transmitted laser power and use a lock-in
amplifier to extract an error signal, which is fed back to the laser.  The 650
nm laser is monitored by a commercial wavelengthmeter (Highfinesse WS7) and
stabilized via feedback from the computer attached to the wavelengthmeter
\cite{Dietrich}.

The trap is loaded using isotope-selective two-step photoionization with a 791
nm ECDL and a N$_2$ laser \cite{Steele2007}.  This allows us to easily switch
between \baseven~and the more common $^{138}\rm{Ba}^{+}$ isotope for use in
various experiments.  Improved reliability and trapping lifetimes have been
observed using this technique when compared to trapping \baseven~without
selective photoionization.  It is not uncommon to hold an ion now for several
days, where previously the lifetime in our trap for \baseven~was measured in
hours.  Our Paul trap is a linear design with 670 volts applied to needle
endcaps, similar to the one described in ref \cite{Olmschenk2007}, driven with
5 watts of RF at 12.39 MHz.  The trap has an axial secular frequency of about
600 kHz and radial frequencies 1.43 and 2.58 MHz.  A DC magnetic field of 8.9 G
created by two current carrying coils is made parallel with the cooling laser
and perpendicular to the optical pumping beam.  This field is necessary to
break degeneracies and prevent the cooling laser from pumping the ion into a
dark state formed by a superposition of Zeeman levels.

The ion can be optically pumped into the $F=2,m_F=0$ state by relying on the
forbidden $F=2,m_F=0\rightarrow F'=2,m_F=0$ transition.  Thus, by setting
the optical pumping beam to have a linear polarization aligned with the
magnetic field, corresponding to $\pi$ polarization, the ion is observed to
optically pump into the $F=2,m_F=0$ state after less than 100 $\mu$s of
exposure time with 93$\pm$1\% fidelity.  Further optimization of the pump laser
polarization is needed to achieve better pumping.

\begin{figure}
\includegraphics[width=4in]{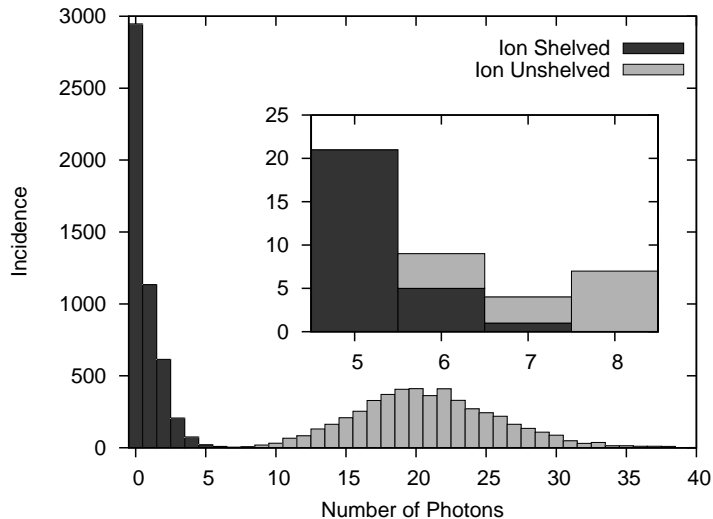}

\caption{A histogram of ion brightness for two cases; where the ion has been
deliberately shelved by disabling the red laser (dark) and where it has not
(light).  Each case has 5000 data points, and represents a cooling laser
exposure time of 10 ms.  The inset shows a zoom in of the overlap region at 6
and 7 photon counts.  In this region, there are 13 offending events out of a
total of 10000 runs, corresponding to a detection fidelity of 99.87\%.}

\label{histogram}
\end{figure}

While the ion is excited into the D$_{5/2}$ state, it is removed from the
cooling cycle and will appear dark when the ion is exposed to the cooling
lasers which normally cause fluorescence \cite{Nagourney1986}.  This
bright/dark signal is the basis of our qubit readout.  Due to the long shelved
state lifetime, it is expected that we can obtain extremely high detection
fidelities, using for example adapative techniques \cite{Myerson2008}.  We
excite the ion directly into the dark state using a 1.76 $\mu$m fiber laser
frequency referenced to a temperature controlled Zerodur\texttrademark~cavity
located in a chamber evacuated to $10^{-7}$~Torr.  The cavity has a free
spectral range of 500 MHz and a finesse of 1000.  With a typical ion brightness
of about 2100 photon counts per second, we obtain better than 99\% bright/dark
detection fidelity in 10 ms, see figure \ref{histogram}.

\begin{figure}
\includegraphics[width=4in]{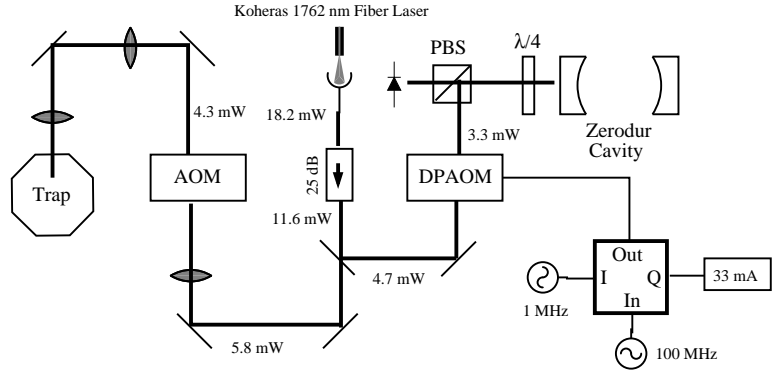}

\caption{The optical setup for the 1.76 $\mu$m laser.  The laser power is split
between laser stabilization and ion state manipulation.  The laser is servo
locked to a Zerodur spaced cavity using a Pound-Drever-Hall technique, where
frequency modulation is introduced by a double-passed AOM (DPAOM), whose drive
frequency is modulated by an I/Q modulator.  The laser light intended for the
ion is switched and frequency shifted by a separate AOM.}

\label{laser-setup}
\end{figure}

The optical setup for the 1.76 $\mu$m laser can be found in figure
\ref{laser-setup}.  Half of the total beam power is diverted to the laser
stabilization setup, while the other half is used to drive the ion.  The beam
used for laser stabilization is double-passed through an AOM which is used both
to shift the laser frequency with respect to the cavity resonance and to
introduce frequency modulation needed for the Pound-Drever-Hall (PDH) lock
\cite{Black2001,Drever1983}.  The beam intended to drive the ion is focused
into an identical AOM, and then two lenses effectively image this focus onto
the ion location.  This configuration allows us to adjust the AOM frequency
without the beam shifting off the ion's location.

To generate the PDH error signal, RF is created by a synthesizer and mixed with
a 1 MHz sine wave and a DC voltage in an I/Q modulator.  This allows high
bandwidth frequency modulation on the output of a synthesized RF signal
\cite{Dietrich}.  This modulation is then mapped onto the 1.76 $\mu$m laser by
application of the double-passed AOM.  With frequency modulation, it is then
possible to observe a PDH error signal from the retroreflected beam of the
Zerodur cavity without resorting to an EOM.

\begin{figure}
%% Generated by script100205-plot
\includegraphics[width=4in]{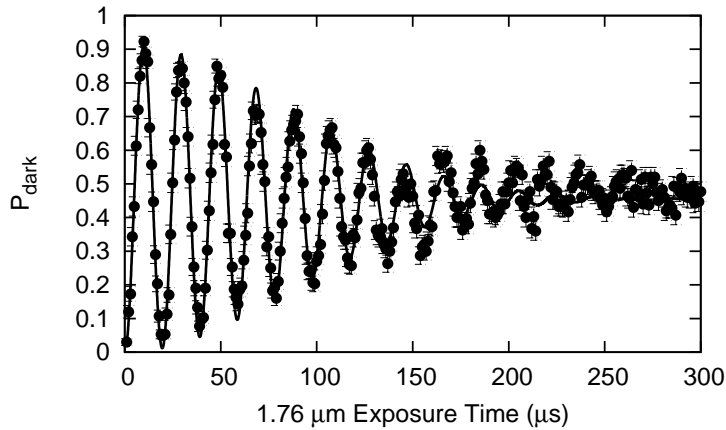}

\caption{Coherent excitation of the 1.76 $\mu$m transition.  The probability of
detecting the ion in the dark state is plotted against the time for which the
laser is exposed to the ion.  Readout is accomplished in about 20 ms by
activating the cooling lasers.  The Rabi frequency is observed to be about 50
kHz with a decay time of $120 \mu {\rm s}$.  These parameters were
determined by a least squares fit to a sine wave with Gaussian decay envelope,
which is shown by the solid line.}

\label{optical-rabi}
\end{figure}

\section{Results}

In a field stronger than about 1 G, the 500 kHz splitting of the $D_{5/2}$ F=3,
F=4 hyperfine levels \cite{Silverans1986} is overcome so that they become an
effective J=4 manifold \cite{Beloy2008}.  In our experiments, it is into the
$J=4,m_J=-2$ level (formed from an $m_F=-1$ and an $m_F=-3$ level) that the
1.76 $\mu$m laser drives the ion from the $F=2,m_F=0$ ground state.  
These two states constitute the levels of our optical qubit.  Coherent optical
nutation on this transition can be seen in figure \ref{optical-rabi}.  The
measured Rabi frequency is about 50 kHz using 4.3 mW of laser power focused to
an intensity of roughly 60 W/cm$^2$, and a coherence time of about 120 $\mu$s
is observed.  In order to obtain this curve, the ion was first optically
pumped, then exposed to 1.76 $\mu$m light for a fixed period of time, and
finally the cooling lasers were enabled and fluorescence detected.  In order to
suppress dephasing due to magnetic field fluctuations, the infrared laser
exposure was triggered on the rising slope of the 60 Hz AC power line voltage.
Although the coherence time is roughly consistent with the laser linewidth,
expected to be about 10 kHz, changes in the coherence time have been observed
by adjusting the ion temperature through manipulation of the cooling laser
frequency.  This suggests that increased coherence might be observed with
improved ion cooling.

\begin{figure}
%% Generated by atlonglast-script
\includegraphics[width=4in]{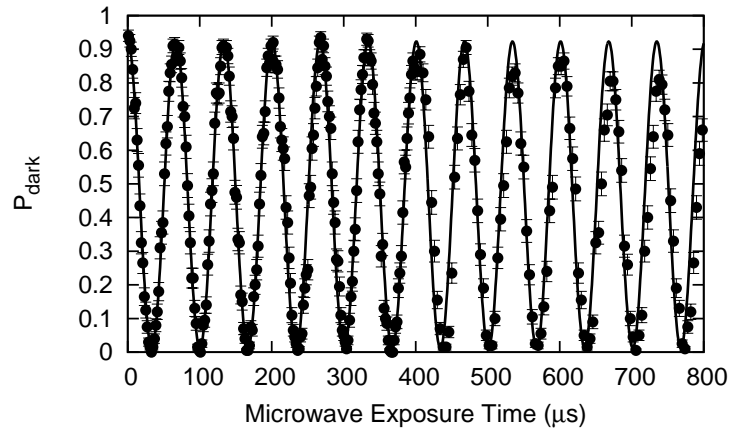}

\caption{The qubit state after a 8 GHz resonant pulse.  The maximum probability
of a dark ion is limited by optical pumping and the shelving efficiency, while
the minimum is limited by coherence of the hyperfine qubit.  At longer exposure
times, the maximum shelving probability is seen to decrease due to magnetic
noise of the 1.76 $\mu$m transition, while the minimum probability remains 0\%.
The Rabi frequency is 15 kHz, again obtained through a fit, which is shown as
a solid line.}

\label{rf-rabi-flops}
\end{figure}

Once a robust readout mechanism is developed, it is possible to use the
hyperfine structure of \baseven~as a qubit.  By exposing the ion to a
$\pi$~pulse of 1.76 $\mu$m light, we can selectively excite the ion from the
$|0\rangle$~qubit state into the shelved state.  An alternative method would be
to use adiabatic passage, which offers more robust population transfer
\cite{Wunderlich2005}.  This technique has also been developed, but is not
implemented here.  Using the magnetically insensitive $m_F=0$ levels of the two
ground state hyperfine levels as our qubit states, coherence times of several
seconds can be achieved \cite{Olmschenk2007}.  Qubit rotations were obtained by
directly exposing the ion to 10 W of resonant microwave radiation at
approximately 8 GHz \cite{Blatt1981,Becker1981} after being optically pumped.
Then, the 1.76 $\mu$m laser is applied to shelve the ion if it is in the higher
energy $F=2,m_F=0$ state, but not in any other hyperfine level, then finally
the cooling lasers are activated and fluorescence detected.  As can be seen in
figure \ref{rf-rabi-flops}, this procedure results in Rabi flops with a
frequency of 15 kHz.  Although the entire sequence was line-triggered , the
increasing length of the microwave pulse caused the 1.76 $\mu$m laser pulse to
occur at a later time in the sequence.  This resulted in the laser frequency
slipping off resonance somewhat, since the pulse began at a different AC phase.
This accounts for the diminishing readout efficiency at longer microwave
exposure times.  Decoherence in the hyperfine qubit itself would appear as an
increasing minimum probability, which is not resolved within 800 $\mu$s.

\section{Conclusion}

Herein we have demonstrated two implementations of ionic qubits using \baseven,
first an optical qubit, and then a hyperfine qubit that takes advantage of the
techniques developed for the optical qubit for readout.  Because of the long
wavelengths involved, it is expected that this advance will facilitate future
developments in ion-photon and long distance ion-ion entanglement, which in
turn is vital for quantum information processing, loop-hole free Bell
inequality tests \cite{Simon2003}, and quantum repeaters for cryptographic 
applications \cite{Duan2001}.

We would like to acknowledge the technical assistance of a number of people,
including Joanna Salacka, Aaron Avril, Ryan Bowler, Adam Kleczewski, Joseph
Pirtle, Chris Dostert, Frank Garcia, Sanghoon Chong, Tom Chartrand, Viki
Mirgon, Eric Magnuson, Peter Greene, Jennifer Porter, Anya Davis, Corey Adams,
and Edan Shahar.  This research was supported by the National Science
Foundation Grants No. 0758025 and No. 0904004, and the Army Research Office
under the DURIP program.


\begin{thebibliography}{29}
\expandafter\ifx\csname natexlab\endcsname\relax\def\natexlab#1{#1}\fi
\expandafter\ifx\csname bibnamefont\endcsname\relax
  \def\bibnamefont#1{#1}\fi
\expandafter\ifx\csname bibfnamefont\endcsname\relax
  \def\bibfnamefont#1{#1}\fi
\expandafter\ifx\csname citenamefont\endcsname\relax
  \def\citenamefont#1{#1}\fi
\expandafter\ifx\csname url\endcsname\relax
  \def\url#1{\texttt{#1}}\fi
\expandafter\ifx\csname urlprefix\endcsname\relax\def\urlprefix{URL }\fi
\providecommand{\bibinfo}[2]{#2}
\providecommand{\eprint}[2][]{\url{#2}}

\bibitem[{\citenamefont{Nielsen and Chuang}(2000)}]{Nielsen}
\bibinfo{author}{\bibfnamefont{M.~A.} \bibnamefont{Nielsen}} \bibnamefont{and}
  \bibinfo{author}{\bibfnamefont{I.~L.} \bibnamefont{Chuang}},
  \emph{\bibinfo{title}{Quantum Computation and Quantum Information}}
  (\bibinfo{publisher}{Cambridge University Press},
  \bibinfo{address}{Cambridge, UK}, \bibinfo{year}{2000}), ISBN
  \bibinfo{isbn}{0-521-63503-9}.

\bibitem[{\citenamefont{Ladd et~al.}(2010)\citenamefont{Ladd, Jelezko,
  Laflamme, Nakamura, Monroe, and O'Brien}}]{Ladd2010}
\bibinfo{author}{\bibfnamefont{T.~D.} \bibnamefont{Ladd}},
  \bibinfo{author}{\bibfnamefont{F.}~\bibnamefont{Jelezko}},
  \bibinfo{author}{\bibfnamefont{R.}~\bibnamefont{Laflamme}},
  \bibinfo{author}{\bibfnamefont{Y.}~\bibnamefont{Nakamura}},
  \bibinfo{author}{\bibfnamefont{C.}~\bibnamefont{Monroe}}, \bibnamefont{and}
  \bibinfo{author}{\bibfnamefont{J.~L.} \bibnamefont{O'Brien}},
  \bibinfo{journal}{Nature} \textbf{\bibinfo{volume}{464}}, \bibinfo{pages}{45}
  (\bibinfo{year}{2010}).

\bibitem[{\citenamefont{{Neuhauser} et~al.}(1980)\citenamefont{{Neuhauser},
  {Hohenstatt}, {Toschek}, and {Dehmelt}}}]{Dehmelt1980}
\bibinfo{author}{\bibfnamefont{W.}~\bibnamefont{{Neuhauser}}},
  \bibinfo{author}{\bibfnamefont{M.}~\bibnamefont{{Hohenstatt}}},
  \bibinfo{author}{\bibfnamefont{P.~E.} \bibnamefont{{Toschek}}},
  \bibnamefont{and}
  \bibinfo{author}{\bibfnamefont{H.}~\bibnamefont{{Dehmelt}}},
  \bibinfo{journal}{Phys. Rev. A.} \textbf{\bibinfo{volume}{22}},
  \bibinfo{pages}{1137} (\bibinfo{year}{1980}).

\bibitem[{\citenamefont{Janik et~al.}(1985)\citenamefont{Janik, Nagourney, and
  Dehmelt}}]{Janik1985}
\bibinfo{author}{\bibfnamefont{G.}~\bibnamefont{Janik}},
  \bibinfo{author}{\bibfnamefont{W.}~\bibnamefont{Nagourney}},
  \bibnamefont{and} \bibinfo{author}{\bibfnamefont{H.}~\bibnamefont{Dehmelt}},
  \bibinfo{journal}{J. Opt. Soc. Am. B} \textbf{\bibinfo{volume}{2}},
  \bibinfo{pages}{1251} (\bibinfo{year}{1985}).

\bibitem[{\citenamefont{DeVoe and Kurtsiefer}(2002)}]{DeVoe2002}
\bibinfo{author}{\bibfnamefont{R.~G.} \bibnamefont{DeVoe}} \bibnamefont{and}
  \bibinfo{author}{\bibfnamefont{C.}~\bibnamefont{Kurtsiefer}},
  \bibinfo{journal}{Phys. Rev. A} \textbf{\bibinfo{volume}{65}},
  \bibinfo{pages}{063407} (\bibinfo{year}{2002}).

\bibitem[{\citenamefont{Dietrich et~al.}(2008)\citenamefont{Dietrich, Avril,
  Bowler, Kurz, Salacka, Shu, and Blinov}}]{Dietrich2008}
\bibinfo{author}{\bibfnamefont{M.~R.} \bibnamefont{Dietrich}},
  \bibinfo{author}{\bibfnamefont{A.}~\bibnamefont{Avril}},
  \bibinfo{author}{\bibfnamefont{R.}~\bibnamefont{Bowler}},
  \bibinfo{author}{\bibfnamefont{N.}~\bibnamefont{Kurz}},
  \bibinfo{author}{\bibfnamefont{J.~S.} \bibnamefont{Salacka}},
  \bibinfo{author}{\bibfnamefont{G.}~\bibnamefont{Shu}}, \bibnamefont{and}
  \bibinfo{author}{\bibfnamefont{B.~B.} \bibnamefont{Blinov}}, in
  \emph{\bibinfo{booktitle}{Non-Neutral Plamsa Physics VII}}, edited by
  \bibinfo{editor}{\bibfnamefont{J.~R.} \bibnamefont{Danielson}}
  \bibnamefont{and} \bibinfo{editor}{\bibfnamefont{T.~S.}
  \bibnamefont{Pedersen}} (\bibinfo{publisher}{AIP Conference Proceedings No.
  1114}, \bibinfo{year}{2008}), pp. \bibinfo{pages}{25--30}.

\bibitem[{\citenamefont{Steele et~al.}(2007)\citenamefont{Steele, Churchill,
  Griffin, and Chapman}}]{Steele2007}
\bibinfo{author}{\bibfnamefont{A.~V.} \bibnamefont{Steele}},
  \bibinfo{author}{\bibfnamefont{L.~R.} \bibnamefont{Churchill}},
  \bibinfo{author}{\bibfnamefont{P.~F.} \bibnamefont{Griffin}},
  \bibnamefont{and} \bibinfo{author}{\bibfnamefont{M.~S.}
  \bibnamefont{Chapman}}, \bibinfo{journal}{Phys. Rev. A}
  \textbf{\bibinfo{volume}{75}}, \bibinfo{eid}{053404} (\bibinfo{year}{2007}).

\bibitem[{\citenamefont{Madej and Sankey}(1990)}]{Madej1990}
\bibinfo{author}{\bibfnamefont{A.~A.} \bibnamefont{Madej}} \bibnamefont{and}
  \bibinfo{author}{\bibfnamefont{J.~D.} \bibnamefont{Sankey}},
  \bibinfo{journal}{Phys. Rev. A} \textbf{\bibinfo{volume}{41}},
  \bibinfo{pages}{2621} (\bibinfo{year}{1990}).

\bibitem[{\citenamefont{Simon and Irvine}(2003)}]{Simon2003}
\bibinfo{author}{\bibfnamefont{C.}~\bibnamefont{Simon}} \bibnamefont{and}
  \bibinfo{author}{\bibfnamefont{W.~T.~M.} \bibnamefont{Irvine}},
  \bibinfo{journal}{Phys. Rev. Lett.} \textbf{\bibinfo{volume}{91}},
  \bibinfo{eid}{110405} (\bibinfo{year}{2003}).

\bibitem[{\citenamefont{Olmschenk et~al.}(2009)\citenamefont{Olmschenk,
  Matsukevich, Maunz, Hayes, Duan, and Monroe}}]{Olmschenk2009}
\bibinfo{author}{\bibfnamefont{S.}~\bibnamefont{Olmschenk}},
  \bibinfo{author}{\bibfnamefont{D.~N.} \bibnamefont{Matsukevich}},
  \bibinfo{author}{\bibfnamefont{P.}~\bibnamefont{Maunz}},
  \bibinfo{author}{\bibfnamefont{D.}~\bibnamefont{Hayes}},
  \bibinfo{author}{\bibfnamefont{L.-M.} \bibnamefont{Duan}}, \bibnamefont{and}
  \bibinfo{author}{\bibfnamefont{C.}~\bibnamefont{Monroe}},
  \bibinfo{journal}{Science} \textbf{\bibinfo{volume}{323}},
  \bibinfo{pages}{486} (\bibinfo{year}{2009}).

\bibitem[{\citenamefont{Moehring et~al.}(2004)\citenamefont{Moehring, Madsen,
  Blinov, and Monroe}}]{Moehring2004}
\bibinfo{author}{\bibfnamefont{D.~L.} \bibnamefont{Moehring}},
  \bibinfo{author}{\bibfnamefont{M.~J.} \bibnamefont{Madsen}},
  \bibinfo{author}{\bibfnamefont{B.~B.} \bibnamefont{Blinov}},
  \bibnamefont{and} \bibinfo{author}{\bibfnamefont{C.}~\bibnamefont{Monroe}},
  \bibinfo{journal}{Phys. Rev. Lett.} \textbf{\bibinfo{volume}{93}},
  \bibinfo{eid}{090410} (\bibinfo{year}{2004}).

\bibitem[{\citenamefont{Gopakumar et~al.}(2002)\citenamefont{Gopakumar,
  Merlitz, Chaudhuri, Das, Mahapatra, and Mukherjee}}]{Gopakumar2002}
\bibinfo{author}{\bibfnamefont{G.}~\bibnamefont{Gopakumar}},
  \bibinfo{author}{\bibfnamefont{H.}~\bibnamefont{Merlitz}},
  \bibinfo{author}{\bibfnamefont{R.~K.} \bibnamefont{Chaudhuri}},
  \bibinfo{author}{\bibfnamefont{B.~P.} \bibnamefont{Das}},
  \bibinfo{author}{\bibfnamefont{U.~S.} \bibnamefont{Mahapatra}},
  \bibnamefont{and}
  \bibinfo{author}{\bibfnamefont{D.}~\bibnamefont{Mukherjee}},
  \bibinfo{journal}{Phys. Rev. A} \textbf{\bibinfo{volume}{66}},
  \bibinfo{pages}{032505} (\bibinfo{year}{2002}).

\bibitem[{\citenamefont{Iskrenova-Tchoukova and
  Safronova}(2008)}]{IskrenovaTchoukova2008}
\bibinfo{author}{\bibfnamefont{E.}~\bibnamefont{Iskrenova-Tchoukova}}
  \bibnamefont{and} \bibinfo{author}{\bibfnamefont{M.~S.}
  \bibnamefont{Safronova}}, \bibinfo{journal}{Phys. Rev. A}
  \textbf{\bibinfo{volume}{78}}, \bibinfo{eid}{012508}
  (\bibinfo{year}{2008}).

\bibitem[{\citenamefont{Sherman
  et~al.}(2005{\natexlab{a}})\citenamefont{Sherman, Trimble, Metz, Nagourney,
  and Fortson}}]{Sherman2005}
\bibinfo{author}{\bibfnamefont{J.~A.} \bibnamefont{Sherman}},
  \bibinfo{author}{\bibfnamefont{W.}~\bibnamefont{Trimble}},
  \bibinfo{author}{\bibfnamefont{S.}~\bibnamefont{Metz}},
  \bibinfo{author}{\bibfnamefont{W.}~\bibnamefont{Nagourney}},
  \bibnamefont{and} \bibinfo{author}{\bibfnamefont{N.}~\bibnamefont{Fortson}},
  in \emph{\bibinfo{booktitle}{2005 Digest of the LEOS Summer Topical
  Meetings}} (\bibinfo{publisher}{IEEE No. 05TH8797},
  \bibinfo{year}{2005}{\natexlab{a}}), \eprint{arXiv:physics/0504013v2}.

\bibitem[{\citenamefont{Madej et~al.}(1992)\citenamefont{Madej, Sankey, Hanes,
  Siemsen, and McKellar}}]{Madej1992}
\bibinfo{author}{\bibfnamefont{A.~A.} \bibnamefont{Madej}},
  \bibinfo{author}{\bibfnamefont{J.~D.} \bibnamefont{Sankey}},
  \bibinfo{author}{\bibfnamefont{G.~R.} \bibnamefont{Hanes}},
  \bibinfo{author}{\bibfnamefont{K.~J.} \bibnamefont{Siemsen}},
  \bibnamefont{and} \bibinfo{author}{\bibfnamefont{A.~R.~W.}
  \bibnamefont{McKellar}}, \bibinfo{journal}{Phys. Rev. A}
  \textbf{\bibinfo{volume}{45}}, \bibinfo{pages}{1742} (\bibinfo{year}{1992}).

\bibitem[{\citenamefont{Fortson}(1993)}]{Fortson1993}
\bibinfo{author}{\bibfnamefont{N.}~\bibnamefont{Fortson}},
  \bibinfo{journal}{Phys. Rev. Lett.} \textbf{\bibinfo{volume}{70}},
  \bibinfo{pages}{2383} (\bibinfo{year}{1993}).

\bibitem[{\citenamefont{Sherman
  et~al.}(2005{\natexlab{b}})\citenamefont{Sherman, Koerber, Markhotok,
  Nagourney, and Fortson}}]{Sherman2005b}
\bibinfo{author}{\bibfnamefont{J.~A.} \bibnamefont{Sherman}},
  \bibinfo{author}{\bibfnamefont{T.~W.} \bibnamefont{Koerber}},
  \bibinfo{author}{\bibfnamefont{A.}~\bibnamefont{Markhotok}},
  \bibinfo{author}{\bibfnamefont{W.}~\bibnamefont{Nagourney}},
  \bibnamefont{and} \bibinfo{author}{\bibfnamefont{E.~N.}
  \bibnamefont{Fortson}}, \bibinfo{journal}{Phys. Rev. Lett.}
  \textbf{\bibinfo{volume}{94}}, \bibinfo{pages}{243001}
  (\bibinfo{year}{2005}{\natexlab{b}}).

\bibitem[{\citenamefont{Dietrich}(2009)}]{Dietrich}
\bibinfo{author}{\bibfnamefont{M.}~\bibnamefont{Dietrich}}, Ph.D. thesis,
  \bibinfo{school}{University of Washington}, \bibinfo{address}{Seattle, WA}
  (\bibinfo{year}{2009}).

\bibitem[{\citenamefont{Olmschenk et~al.}(2007)\citenamefont{Olmschenk, Younge,
  Moehring, Matsukevich, Maunz, and Monroe}}]{Olmschenk2007}
\bibinfo{author}{\bibfnamefont{S.}~\bibnamefont{Olmschenk}},
  \bibinfo{author}{\bibfnamefont{K.~C.} \bibnamefont{Younge}},
  \bibinfo{author}{\bibfnamefont{D.~L.} \bibnamefont{Moehring}},
  \bibinfo{author}{\bibfnamefont{D.}~\bibnamefont{Matsukevich}},
  \bibinfo{author}{\bibfnamefont{P.}~\bibnamefont{Maunz}}, \bibnamefont{and}
  \bibinfo{author}{\bibfnamefont{C.}~\bibnamefont{Monroe}},
  \bibinfo{journal}{Phys. Rev. A} \textbf{\bibinfo{volume}{76}},
  \bibinfo{pages}{052314} (\bibinfo{year}{2007}).

\bibitem[{\citenamefont{Nagourney et~al.}(1986)\citenamefont{Nagourney,
  Sandberg, and Dehmelt}}]{Nagourney1986}
\bibinfo{author}{\bibfnamefont{W.}~\bibnamefont{Nagourney}},
  \bibinfo{author}{\bibfnamefont{J.}~\bibnamefont{Sandberg}}, \bibnamefont{and}
  \bibinfo{author}{\bibfnamefont{H.}~\bibnamefont{Dehmelt}},
  \bibinfo{journal}{Phys. Rev. Lett.} \textbf{\bibinfo{volume}{56}},
  \bibinfo{pages}{2797} (\bibinfo{year}{1986}).

\bibitem[{\citenamefont{Myerson et~al.}(2008)\citenamefont{Myerson, Szwer,
  Webster, Allcock, Curtis, Imreh, Sherman, Stacey, Steane, and
  Lucas}}]{Myerson2008}
\bibinfo{author}{\bibfnamefont{A.~H.} \bibnamefont{Myerson}},
  \bibinfo{author}{\bibfnamefont{D.~J.} \bibnamefont{Szwer}},
  \bibinfo{author}{\bibfnamefont{S.~C.} \bibnamefont{Webster}},
  \bibinfo{author}{\bibfnamefont{D.~T.~C.} \bibnamefont{Allcock}},
  \bibinfo{author}{\bibfnamefont{M.~J.} \bibnamefont{Curtis}},
  \bibinfo{author}{\bibfnamefont{G.}~\bibnamefont{Imreh}},
  \bibinfo{author}{\bibfnamefont{J.~A.} \bibnamefont{Sherman}},
  \bibinfo{author}{\bibfnamefont{D.~N.} \bibnamefont{Stacey}},
  \bibinfo{author}{\bibfnamefont{A.~M.} \bibnamefont{Steane}},
  \bibnamefont{and} \bibinfo{author}{\bibfnamefont{D.~M.} \bibnamefont{Lucas}},
  \bibinfo{journal}{Phys. Rev. Lett.} \textbf{\bibinfo{volume}{100}},
  \bibinfo{eid}{200502} (\bibinfo{year}{2008}).

\bibitem[{\citenamefont{Black}(2001)}]{Black2001}
\bibinfo{author}{\bibfnamefont{E.~D.} \bibnamefont{Black}},
  \bibinfo{journal}{Am. J. Phys.} \textbf{\bibinfo{volume}{69}},
  \bibinfo{pages}{79} (\bibinfo{year}{2001}).

\bibitem[{\citenamefont{Drever et~al.}(1983)\citenamefont{Drever, Hall,
  Kowalski, Hough, Ford, Munley, and Ward}}]{Drever1983}
\bibinfo{author}{\bibfnamefont{R.~W.~P.} \bibnamefont{Drever}},
  \bibinfo{author}{\bibfnamefont{J.~L.} \bibnamefont{Hall}},
  \bibinfo{author}{\bibfnamefont{F.~V.} \bibnamefont{Kowalski}},
  \bibinfo{author}{\bibfnamefont{J.}~\bibnamefont{Hough}},
  \bibinfo{author}{\bibfnamefont{G.~M.} \bibnamefont{Ford}},
  \bibinfo{author}{\bibfnamefont{A.~J.} \bibnamefont{Munley}},
  \bibnamefont{and} \bibinfo{author}{\bibfnamefont{H.}~\bibnamefont{Ward}},
  \bibinfo{journal}{Appl. Phys. B: Lasers and Optics}
  \textbf{\bibinfo{volume}{31}}, \bibinfo{pages}{97} (\bibinfo{year}{1983}).

\bibitem[{\citenamefont{Silverans et~al.}(1986)\citenamefont{Silverans, Borghs,
  De~Bisschop, and Van~Hove}}]{Silverans1986}
\bibinfo{author}{\bibfnamefont{R.~E.} \bibnamefont{Silverans}},
  \bibinfo{author}{\bibfnamefont{G.}~\bibnamefont{Borghs}},
  \bibinfo{author}{\bibfnamefont{P.}~\bibnamefont{De~Bisschop}},
  \bibnamefont{and} \bibinfo{author}{\bibfnamefont{M.}~\bibnamefont{Van~Hove}},
  \bibinfo{journal}{Phys. Rev. A} \textbf{\bibinfo{volume}{33}},
  \bibinfo{pages}{2117} (\bibinfo{year}{1986}).

\bibitem[{\citenamefont{Beloy et~al.}(2008)\citenamefont{Beloy, Derevianko,
  Dzuba, Howell, Blinov, and Fortson}}]{Beloy2008}
\bibinfo{author}{\bibfnamefont{K.}~\bibnamefont{Beloy}},
  \bibinfo{author}{\bibfnamefont{A.}~\bibnamefont{Derevianko}},
  \bibinfo{author}{\bibfnamefont{V.~A.} \bibnamefont{Dzuba}},
  \bibinfo{author}{\bibfnamefont{G.~T.} \bibnamefont{Howell}},
  \bibinfo{author}{\bibfnamefont{B.~B.} \bibnamefont{Blinov}},
  \bibnamefont{and} \bibinfo{author}{\bibfnamefont{E.~N.}
  \bibnamefont{Fortson}}, \bibinfo{journal}{Phys. Rev. A}
  \textbf{\bibinfo{volume}{77}}, \bibinfo{pages}{052503}
  (\bibinfo{year}{2008}).

\bibitem[{\citenamefont{Wunderlich et~al.}(2007)\citenamefont{Wunderlich,
  Hannemann, K\"orber, H\"affner, Roos, H\"ansel, Blatt, and
  Schmidt-Kaler}}]{Wunderlich2005}
\bibinfo{author}{\bibfnamefont{C.}~\bibnamefont{Wunderlich}},
  \bibinfo{author}{\bibfnamefont{T.}~\bibnamefont{Hannemann}},
  \bibinfo{author}{\bibfnamefont{T.}~\bibnamefont{K\"orber}},
  \bibinfo{author}{\bibfnamefont{H.}~\bibnamefont{H\"affner}},
  \bibinfo{author}{\bibfnamefont{C.}~\bibnamefont{Roos}},
  \bibinfo{author}{\bibfnamefont{W.}~\bibnamefont{H\"ansel}},
  \bibinfo{author}{\bibfnamefont{R.}~\bibnamefont{Blatt}}, \bibnamefont{and}
  \bibinfo{author}{\bibfnamefont{F.}~\bibnamefont{Schmidt-Kaler}},
  \bibinfo{journal}{J. Mod. Optics} \textbf{\bibinfo{volume}{54}},
  \bibinfo{pages}{1541} (\bibinfo{year}{2007}), \eprint{quant-ph/0508159}.

\bibitem[{\citenamefont{Blatt and Werth}(1982)}]{Blatt1981}
\bibinfo{author}{\bibfnamefont{R.}~\bibnamefont{Blatt}} \bibnamefont{and}
  \bibinfo{author}{\bibfnamefont{G.}~\bibnamefont{Werth}},
  \bibinfo{journal}{Phys. Rev. A} \textbf{\bibinfo{volume}{25}},
  \bibinfo{pages}{1476} (\bibinfo{year}{1982}).

\bibitem[{\citenamefont{Becker et~al.}(1981)\citenamefont{Becker, Blatt, and
  G.~Werth}}]{Becker1981}
\bibinfo{author}{\bibfnamefont{W.}~\bibnamefont{Becker}},
  \bibinfo{author}{\bibfnamefont{R.}~\bibnamefont{Blatt}}, \bibnamefont{and}
  \bibinfo{author}{\bibfnamefont{G.}~\bibnamefont{G.~Werth}},
  \bibinfo{journal}{J. Phys. Colloques} \textbf{\bibinfo{volume}{42}},
  \bibinfo{pages}{339} (\bibinfo{year}{1981}).

\bibitem[{\citenamefont{Duan et~al.}(2001)\citenamefont{Duan, Lukin, Cirac, and
  Zoller}}]{Duan2001}
\bibinfo{author}{\bibfnamefont{L.-M.} \bibnamefont{Duan}},
  \bibinfo{author}{\bibfnamefont{M.~D.} \bibnamefont{Lukin}},
  \bibinfo{author}{\bibfnamefont{J.~I.} \bibnamefont{Cirac}}, \bibnamefont{and}
  \bibinfo{author}{\bibfnamefont{P.}~\bibnamefont{Zoller}},
  \bibinfo{journal}{Nature} \textbf{\bibinfo{volume}{414}},
  \bibinfo{pages}{413} (\bibinfo{year}{2001}).

\end{thebibliography}
\end{document}